\documentstyle[preprint,aps]{revtex}

\begin{document}

\thispagestyle{empty}

\title{Rho-Omega Mixing Off Shell and Charge  Symmetry Breaking
In the N-N
Potential }

\author{Thomas D. Cohen$^{a,b}$\thanks{Permanent address:
Department of Physics and Center for Theoretical Physics,
University of Maryland, College Park, MD 20742, USA},
 and Gerald A. Miller$^a$}

\address{Department of Physics$^a$
FM-15 and
Institute For Nuclear Theory$^b$
NK-12, University of Washington,
Seattle, Washington 98195, USA}

\maketitle

\begin{abstract}
The role of the off-shell dependence of $\rho-\omega$ mixing in
the
charge symmetry breaking nucleon-nucleon potential is discussed.
It is
shown that models
describing the off-shell dependence of $\rho-\omega$ mixing are
not
sufficient to determine the
charge symmetry breaking nucleon-nucleon potential.
\end{abstract}
\newpage
\section{Introduction}

Charge symmetry breaking has been studied for a long time,
see  e.g. the
reviews\cite{NS69,H69,SHLOMO,HM79,MNS90,NMS92,MVO94}
and especially the references therein. We follow
ref.~\cite{MVO94}
in  summarizing  a few main features.
Charge independence and charge  symmetry breaking is caused by
the $d$-$u$
quark mass difference $m_d-m_u >0$, and electromagnetic effects.
The general goal of this area of research is to find small but
observable effects of the breaking  of charge  independence and
charge
symmetry.   This
provides significant insight into strong interaction dynamics
since the
underlying origin of the breaking is understood.  Over the years
there
has  been substantial
experimental and theoretical progress. First,
we recall the old idea that $m_d-m_u>0$ along with
electromagnetic
effects accounts for  the observed mass
differences between members of hadronic isospin  multiplets.
This mass difference also leads to the notion that the physical
$\rho$ and
$\omega$ mesons are
isospin mixed superpositions of bare states of good isospin.
Indeed,
substantial effects
of $\rho-\omega$ mixing have been observed in the $e^+e^-
\rightarrow \pi^+\pi^-$ cross section at $q^2\approx m^2_\omega$
\cite{Q78,B85}.
 These results
allow an extraction of the strong contribution to the
$\rho-\omega$ mixing
matrix element $<\rho|H_{str}|\omega>\approx$ - 5200~MeV$^2$.
\cite{M94,MVO94}
Two nucleons
may exchange a mixed $\rho-\omega$ meson. If one uses
$<\rho|H_{str}|\omega>\approx$ - 5200~MeV$^2$ one obtains a
nucleon-nucleon
interaction which is consistent with  the experimental value
$\Delta a_{CSB} = a^N_{pp}-a^N_{nn} = 1.5\pm
0.5$ fm \cite{CB87}.
 Such a force can also consistently account most of the strong
interaction
contribution to the $^3H$-$^3He$ binding energy difference
\cite{CB87} and for
much of the Nolen-Schiffer anomaly\cite{BI87}.
The TRIUMF (477~MeV \cite{A86} and 350~MeV\cite{GV369}) and IUCF
(183~MeV)
\cite{K90}
experiments  have compared
analyzing powers of $\vec np$ and $n\vec p$ scattering and
observe charge
symmetry breaking at the
level expected from $\pi,\gamma$ and $\rho$-$\omega$ exchange
effects. The
latter effects are important at 183~MeV.
Thus $\rho$-$\omega$ mixing seems to
describe most of the observed features
or charge symmetry breaking in nuclear physics.   While it is
certainly true that other  mechanism cannot be ruled out,
$\rho$-$\omega$ mixing appears to give a consistent description
of the bulk of the experimental data.

Recently, this success has been called into question.
The
momentum dependence of the $\rho-\omega$ mixing amplitude has
been
calculated using several
different models\cite{GHT92,PW92,KTW93,MTRC94}.
While these models are
based on quite different physical assumptions, they all share
one important quality: the $\rho-\omega$ mixing at spacelike
momenta
in all of these models is quite different from
 its value at the $\omega$ pole---generally of the opposite
sign and significantly reduced in magnitude.  Indeed it has been
shown
that for a wide
class of models\cite{OCPTW94} the mixing must go to zero at
$q^2=0$ implying
that amplitude changes sign.  Moreover, a
QCD sum-rule calculation,  also apparently gives a  similarly
large momentum dependence of the coupling\cite{HHKM94}.
  Since the N-N
potential probes  the spacelike region, this appears to imply
that the
vector meson exchange part of the charge symmetry
breaking nucleon-nucleon NN potential is very different from one
based on
the on-mass-shell mixing.  Indeed,  NN potentials have been
constructed based on these momentum dependent mixing amplitudes
and
these are quite different from the ones used
in the successful
phenomenology\cite{GHT92}-\cite{MTRC94},\cite{IN94}.

The purpose of the present paper is to study the
general role of the off-shell meson propagator in
NN potentials. We find that
knowledge of the off-shell meson propagator is not sufficient to
determine the potential. In particular, one needs the vertex
functions computed from the same theory that supplied the
propagator. None of the present treatments of the off-shell
propagator deals with the issue of the necessary vertex
functions. It is not our intent to compute these functions.
Rather, we wish to clarify issues of principle. Accordingly we
have included
a number of simple illustrative examples.
We do show, however, that the CSB induced by the $\rho-\omega$
exchange potential can account for the existing data even if the
the $q^2$ dependence is exactly as specified in any of the
references
\cite{GHT92}-\cite{MTRC94},\cite{IN94},\cite{HHKM94}. This is
done by using CSB vertex functions

We turn to an outline of this paper.
In Sect. II we discuss the problem
that in hadronic field theoretic models
there is never
a unique choice for fields, even in a renormalizable
theory\cite{Haag},
\cite{Ruelle},  \cite{Borchers} and \cite{Coleman}.
 This means that the propagator and the vertex functions are  not
unique.
 We argue generally and with
two explicit examples that while the
propagator depends on the choice of field variables, the
observables do not.
Thus, knowledge of the off-shell meson
propagator by itself gives
no information unless one knows which field is used.
One may be able to
deduce which definition of the  field has been used from a
complete theory by studying the interactions with the other
degrees
of freedom in the problem.  However, if the
theory is incomplete and the interactions
of the field with all of the other degrees of freedom is unknown,
knowledge of the off-shell
propagator by itself is not physically meaningful.

There is
an even more serious problem.  Modern
meson exchange potentials are motivated by
field theoretic concerns.  However,
there is  no first principle method for obtaining the ``correct''
NN potential directly from either QCD, or from some hadron
field theoretic model or from any experimentally
accessible set of data of hadronic properties.
 Given this essential
difficulty, we believe it is sensible to
adopt the general approach used in the construction
of meson exchange potentials to the case
of charge symmetry breaking.  This approach makes the
pragmatic assumptions of including the long range features
in the meson propagators and the short range
features in the vertex functions. This separation is discussed in
Sect. III. Such a separation
may be questioned, but {\it a
priori} these assumptions should be no worse for the case of  CSB
potentials then they are for the isospin
conserving part of the interaction.  Moreover given the  lack of
rigor in the construction of potential from the underlying field
theory, some  assumptions must be made in order to make  make any
connection between  $\rho-\omega$ mixing and the CSB potential.
Given this, it is highly desirable to make sure that the
assumptions are consistent with those made elsewhere in the
problem.

It is worth stressing at the outset, that in
conventional  treatments of meson exchange potentials the
off-shell propagator plays no role.  This is discussed in Sect.
IV where realistic boson-exchange  charge symmetric
potentials are defined to be those that are consistent with the
separation discussed in Sect.III. We show that for models with
realistic spectral functions the momentum dependence of the
meson propagator can be absorbed into that of the vertex
function. An example of an unrealistic momentum dependent
$\omega$ self energy is presented.

The ideas of the Sects. II-IV are applied to the CSB potential
caused by $\rho-\omega $ exchange in Sect. V. We show that the
influence
of the momentum dependence of the $\rho-\omega $ mixing matrix
element
can be included by allowing the $\rho$-nucleon coupling constant
to violate charge symmetry. In particular, if the model
of Ref.\cite{HHKM94} is used
one needs  CSB coupling constants that are 0.8\% of the standard
coupling
constants to reproduce the
results of a potential obtained without momentum dependence in
the
$\rho-\omega $ mixing matrix element and without CSB in the
coupling constants.
We summarize the
analysis in Sect. VI.

\section{Field Redefinitions And Off-Shell Propagators
\label{fr}}

It has been known for quite some time that value of an off shell
propagator
is completely dependent on the choice of field.
This is an example of a general theorem proved by
Haag\cite{Haag},
Ruelle\cite{Ruelle}, and Borchers\cite{Borchers} which has been
discussed by Coleman, Wess and Zumino\cite{Coleman}.
The off-shell propagators depend on the choice of
interpolating fields, whereas all S-matrix elements are
independent of this choice. Thus an off-shell propagator,
taken in isolation, can have no physical meaning.

To illustrate why this is so, let us
consider the simplest possible case, the field corresponding to a
stable scalar particle
in some nontrivial interacting field theory.  The equation of
motion for this
system may be written as
\begin{equation}
\Box \phi(x) + m^2 \phi(x) = -j(x).
\end{equation}
This equation of motion is determined from a Lagrangean density
${\cal L}(\phi,j)$.
Furthermore, let us insist on studying the renormalized field,
mass  and current. This
means that the correlation function for $\phi$ will have a pole
with residue of unity at the physical mass, $m$:
\begin{equation}
\langle \phi, \vec{p} | \phi(x) | {\rm vac} \rangle  =
e^{ip\cdot x}
\end{equation}
which implies that
\begin{equation}
\lim_{q^2 \rightarrow m^2}\, (q^2 - m^2) \,  \int {\rm d}^4 x \,
e^{i q \cdot x}
\, \langle {\rm vac}| {\rm T}[\phi(x) \phi(0)] |{\rm vac} \rangle
\,
= \, i .
\label{renorm}
\end{equation}

We are concentrating on
the renormalized
quantities because  un-renormalized properties are not observable
and depend on the details of the renormalization procedure.
Ultimately we will be interested  in the spectral decomposition
of the propagator in terms of the physical states of the
system and this is directly related to the {\it renormalized}
fields and sources.

It is worth noting that the renormalization conditions put quite
stringent constraints on  matrix elements of the renormalized
source.  In particular they imply that source does not connect
the vacuum to a one particle
state
\begin{equation}
\langle \phi , \vec{p} | j(0) | {\rm vac} \rangle \, = \, 0 .
\label{cond}
\end{equation}
This can be seen simply:
$$ \langle \phi , \vec{p} | j(x) | {\rm  vac}\rangle = \langle
\phi ,
\vec{p} | (\Box + m^2) \phi(x) | {\rm vac}\rangle  = (-p^2 + m^2)
\langle
\phi , \vec{p} | \phi(x) | {\rm vac} \rangle
$$
where $p^2$ is the square of the four momentum of the state which
is $m^2$.

Now we come to the crux of the issue.  There is enormous freedom
in the
choice of field variables, and consequently the Green's
functions. In particular, we introduce a new renormalized field
and a new source current according to:
\begin{eqnarray}
\phi^{\prime}(x) & = & \phi(x) + a(x), \label{def}\\
j^{\prime}(x)& = & j(x) + (\Box + m^2)\,     a(x) \label{jpdef}
\end{eqnarray}
where $a(x)$ is an
operator such that
$\langle {\rm vac} | a| \phi, \vec{p} \rangle = 0$.
 Thus, for example, $a(x)$ may be
a multiple of the renormalized source $j(x)$ or $a(x)$ could have
the form $a(x) = (\Box + m^2) b(x)$
where $b(x)$ is an arbitrary renormalized local composite
operator.
The new field and source satisfies an equation of motion with the
same form as
the original:
\begin{equation}
(\Box + m^2) \phi^{\prime}(x) = j^{\prime}(x) \label{eomp}
\end{equation}
It also satisfies the same renormalization conditions
\begin{eqnarray}
\langle \phi, \vec{p} | \phi^{\prime}(x) | {\rm vac} \rangle & =
&
e^{i p\cdot x}
\\
\lim_{q^2 \rightarrow m^2} \, (q^2 - m^2) \,  \int {\rm d}^4 x \,
e^{i q \cdot x}
\, \langle {\rm vac}|{\rm T}[\phi^{\prime}(x)
\phi^{\prime}(0)]|{\rm vac} \rangle \,
& = &\, i  \label {renorm2}
\end{eqnarray}

The field variable $\phi^{\prime}$ is as good a choice for the
field variable
as the original  field $\phi$---its equation of motion is of the
same form and it satisfies the correct renormalization
conditions.
It makes no difference to any
{\it physical}
amplitude whether one chooses to describe the physics  in terms
of the field  $\phi$ or $\phi^{\prime}$.  Thus, the masses  of
particles and possible  bound states and S matrix  elements  for
scattering states must be identical with either description.
Going from one to the other amounts to nothing more
than a change of variables.

While the physics clearly does not depend on which field is
chosen, the propagator depends strongly on this choice:
\begin{eqnarray}
\int {\rm d}^4 x \, e^{i q \cdot x}
\, \langle {\rm vac}| {\rm T}[\phi^{\prime}(x) \phi^{\prime}(0)]
|{\rm vac}\rangle \, \nonumber\\
= \int {\rm d}^4 x \, e^{i q \cdot x}
\, \langle {\rm vac}|{\rm T}[\phi(x) \phi(0)] \rangle +\int {\rm
d}^4 x \,
e^{i q \cdot x}
\, \langle {\rm T}[\phi(x) a(0)] + {\rm T}[a(x) \phi(0)] + {\rm
T}[a(x) a(0)]|{\rm vac}\rangle
\end{eqnarray}
Eq.(\ref{renorm2}), which picks out the pole
 at $q^2 = m^2$, is obtained
since by construction $a$ does not connect the vacuum to the one
$\phi$ state.  Clearly, this is necessary since the correlation
functions for $\phi$ and
$\phi^{\prime}$ satisfy the renormalization conditions in eqs.
(\ref{renorm}) and (\ref{renorm2}).
Off shell, however, there is
no requirement that this term vanish and the two propagators will
in general differ.  Moreover, since the overall scale of $a$ is
arbitrary it is clear that one can make the difference between
the two descriptions arbitrarily large.

Let us make these ideas explicit by  considering  two examples
from a theory in which the current
$j$ is a static external source.
In this case the energy of the system is given by
\begin{eqnarray}
E=\int {\rm d}^3r {1\over 2} j(\vec r)\phi(\vec r)
\end{eqnarray}
or
\begin{eqnarray}
E=\int {\rm d}^3r {1\over 2} j(\vec r) G(\vec r,\vec
r^{\,\prime})
j(\vec r^{\,\prime}),
\end{eqnarray}
where $G(\vec r,\vec r^{\,\prime})$
is the inverse of the operator $\nabla^2-m^2$.
Let us first take $a(x)$ to be a simple function of $\vec x$,
which is
independent of $\phi$. Then $\langle \phi,\vec p|a(x)|{\rm
vac}\rangle=0$
 and the
renormalization conditions of eqns. (\ref{renorm},\ref{renorm2})
are satisfied. One may determine a new Lagrangean density ${\cal
L}^\prime$
and a new Hamiltonian ${\cal H}^\prime$ by starting with the
original
${\cal L}$ and transforming the variables. Then the new energy
$E^\prime$
is given by
\begin{eqnarray}
E^\prime=\int {\rm d}^3r {1\over 2} [\vec \nabla
(\phi^\prime(\vec r)+a(\vec r)\cdot(\phi^\prime(\vec r)+a(\vec
r))
+ \nonumber\\
m^2(\phi^\prime(\vec r)+a(\vec r))^2 +2
j(\vec r)(\phi^\prime(\vec r) +a(\vec r))],
\end{eqnarray}
and using the equation of motion (\ref{eomp}) in the static
limit leads to
\begin{eqnarray}
E^\prime=\int {\rm d}^3r {1\over 2}j(\vec r)(\phi^\prime(\vec r)
+a(\vec r)).
\end{eqnarray}
But Eq.(\ref{def}) tells us that $E^\prime=E$.
Even though the current $j^\prime$ of  Eq.(\ref{jpdef}) is
different
than $j$ the energy of the system does not depend on the
choice of the function $a(\vec x)$.

A more interesting example is obtained by letting
$\phi=(1+f(\vec x))\phi^\prime $
(or $a(\vec x)= -{f(\vec x)\over 1+f(\vec x)} \phi(x)).$
We place the static source j
at the origin and choose $f(\vec x)$ to vanish at large
values of $|\vec x|$ faster than $e^{-m|\vec x|}/|\vec x|$.
This maintains the original value of
the renormalized
coupling constant ( which is proportional to the asymptotic
field) and therefore is the analog of our renormalization for
problems with static sources.
In this case the equation of
motion is
\begin{equation}
D\phi^\prime=-j^\prime
\label {neom}
\end{equation}
where
\begin{equation}
D\equiv
(1+f)^2 (-\nabla^2+m^2) -2(1+f)
\partial_\mu f \partial^\mu -(1+f)\nabla^2f
\end{equation}
and
\begin{equation}
j^\prime\equiv
j(1+f).
\end{equation}
Clearly the Green's function $G^\prime$ ( the inverse of  $D$)
and current
$j^\prime$ are both fairly
complicated. The use of the new Hamiltonian density ${\cal
H}^\prime $ gives
\begin{eqnarray}
E^\prime=\int {\rm d}^3 r{1\over 2}[ (1+f)^2
\vec\nabla\phi^\prime\cdot\vec\nabla\phi^\prime \nonumber \\
+(\phi^\prime)^2 \vec\nabla f\cdot\vec\nabla f
+2\phi^\prime(1+f)\vec\nabla f\cdot\vec\nabla\phi^\prime+
(m\phi^\prime)^2(1+f)^2+2j\phi^\prime(1+f)].
\end{eqnarray}
Integration by parts  and
the equation of motion Eq.(\ref{neom}) allows one to obtain
\begin{equation}
E^\prime \, = \, {1\over 2}\int {\rm d}^3 r
j(1+f)\phi^\prime \, =\,{1\over 2}\int {\rm d}^3 r j^\prime(\vec
r)G^\prime(\vec
r,\vec r^\prime)j^\prime(\vec r^\prime),
\end{equation}
which is just the original energy since $(1+f)\phi^\prime=\phi$.

Thus we have seen two explicit examples in which
transformations of field variables change the equation of
motion, the Green's functions and the currents without
changing, the physical observable, the energy of the system.

These same arguments of Eqns. (5-10) can be used
to show that the various
n-point vertex functions of the also depend  on the
specific choice of field.
The generalization of the argument to vector fields rather than
scalars and to correlation functions of two different fields
uses standard techniques.
Again one finds that off-shell propagators
and the vertex functions depend explicitly on the choice of
field.

It is clear what is going on here. Neither the off-shell
propagators nor the vertex functions are directly observable.
{}From a
theoretical  point of view, the values of these quantities depend
explicitly on which arbitrary choice of field one  makes.
Various combinations
of the propagators and the vertex functions correspond to
physical quantities and it is only these combinations which can
be measured.  Choosing a particular field amounts to making a
bookkeeping choice---it only determines whether some bit of the
physics will be found in the vertex or in the propagator.

The point we wish to stress is that knowledge, however precise,
of
the off-shell propagator contains no
physical information unless one specifies
the choice of the quantum field or equivalently unless one has
knowledge of
how the field couples to the rest of the system---{\it i.e.}
knowledge of  the vertex functions which arise from the
same field choice.  Thus, a model for the
off-shell propagator in the absence of a
{\it consistent} model for the vertex
functions is not complete.
The models of refs.
\cite{GHT92}-\cite{MTRC94},\cite{IN94} present
the mixed $\rho-\omega$ propagator off-shell, but
do not give the necessary
simultaneous  consistent description of the
CSB N-N-vector meson vertices.

\section{Philosophy Of Meson Exchange Potentials}

The preceeding argument that off-shell meson propagators are not
sufficient
is entirely based on
field theoretic considerations.  Clearly, this does not help us
to
to compute
observables, it does not address the
question of how one can compute CSB (or any other) observables in
nuclear
physics. One typically
constructs a nucleon-nucleon potential and then computes
wavefunctions,
hoping
that the potentials capture the essential aspects
of the underlying field theory.  Nevertheless,
there is no unambiguous way to construct
potentials. Nontrivial assumptions must be made.

Here we will assume that the assumptions underlying
phenomenologically successful  meson
exchange models are reasonable.
While
one can construct equally successful purely phenomenological
models, the meson exchange models
make a connection to the spectral
properties of the underlying theory.  Moreover, the entire
question we are investigating---the role
of $\rho-\omega$ mixing in CSB effects in nuclear
physics---can only be addressed in the context of a
potential model which employs vector mesons.

There is  a definite philosophy underlying the construction of
NN potentials from  meson exchange.
One principal idea is the need for
 a separation of momentum  or length  scales.   One  explicitly
includes the exchange of mesons lighter  (and hence more
long-ranged) than
some scale separation point.  All short ranged
effects are either incorporated in
phenomenologically determined  vertex functions
or by some other purely phenomenological
means.  The physical picture underlying this philosophy is that
the nucleon  has a  three-quark  core  which cannot be described
efficiently in
terms of mesons, while at longer distances the
nucleon structure is dominated by a meson cloud.

 To some extent, the fact that short ranged effects
are handled as pure phenomenology is of little importance in most
low energy nuclear physics applications. Because of repulsion at
short distances, nuclear wave functions have strong short
distance
correlations which prevent the system from feeling the truly
short
range part of the potential.  Moreover, at very short distances
the
concept of an NN potential becomes particularly inappropriate.
Typically, in meson exchange potentials this scale separation
point,
which we will call $\Lambda_s$, is taken to be of order 1 GeV so
that
$\rho$ and $\omega$ mesons
are explicitly included while heavier vector mesons are not.  It
is worth
observing, however, that this does not mean that the short
distance
physics does not have important long range consequences.  In
particular,
the value of the meson-nucleon coupling constant,
determined by
short distanced physics,
plays an essential role in the potentials at long and
intermediate
ranges.

We believe that this general approach of treating the short range
part
of the NN interaction  phenomenologically while explicitly
including the
effects of lighter mesons is reasonable.  This
general approach ought to be applicable to charge symmetry
breaking effects.

There is another important assumption which underlies these
models.   It is
 assumed that at except at short
distances the vector part of the potential is
dominated by the vector mesons.  Thus it is assumed that
continuum two pion vector-isovector and three pion
vector-isoscalar exchange contributions are small--- {\it i.e.}
that the only substantial strength arising from the two pion
vector-isovector exchange is
sufficiently concentrated at the $\rho$
mass as to be well described by  $\rho$ exchange and analogously
for three pions and the $\omega$ exchange.  We note that this
assumption can be questioned. In its
favor we note that in $e^+ e^- \rightarrow$ pions, the $\rho$
and $\omega$ peaks do, in fact, completely dominate the low lying
spectral function.

In our discussions we will adopt the Bonn potential \cite{BONN}
strategy of incorporating all short range effects in vertex
functions.
In such a strategy the scale separation between long and short
range
is particularly easy to enforce: the phenomenological vertex
functions
are analytic for  $q^2 < \Lambda_s^2$ while the propagators
are analytic for $q^2 > \Lambda_s^2$, where $q^2$ is the square
of the
four momentum.

\section{Momentum-Dependent Self-Energies in
Meson-Exchange Potentials \label{csc} }

It is probably useful to discuss  an analogous, and perhaps
somewhat simpler problem
before discussing charge symmetry
breaking.   The $\rho-\omega$ mixing matrix element
is an off-diagonal mass term.  Models which give momentum
dependence to
this off-diagonal mass can also be expected to give momentum
dependence to the analogous diagonal mass terms---{\it i.e.}
to the vector meson self-energies.
It is clearly useful to understand
the role of the
momentum
dependence of the $\rho$ and $\omega$ self energies in the charge
symmetry preserving potential before taking on the challenge
of understanding the
the momentum dependence of the  $\rho-\omega$ mixing.

For simplicity we examine the one    $\omega$ exchange
contribution.   First consider the traditional meson exchange
model description with the  scale separation   as outlined above.
 The
potential is given by
\begin{equation}
V_{\omega} (q^2) = \frac{( g_\omega^{ \rm
v}(q^2)\gamma^{(1)}_{\mu} \,
+  \, g_\omega^{ \rm t}(q^2)   q^\alpha \sigma^{(1)}_{\alpha \mu}
 )
(g^{\mu \nu} - q^{\mu}q^{\nu}/m_{\omega}^2) (g_{\omega}^{\rm
v}(q^2)
\gamma^{(2)}_{\nu} + g_{\omega}^{\rm t}(q^2) \sigma_{\nu \beta}
q^\beta
)  }{q^2 - m_\omega^2}
 \label{oep}
\end{equation}
where $g_{\omega}^{\rm v} (q^2)$ and $g_{\omega}^{\rm t}(q^2)$
are the
vector and tensor couplings of the omega to the nucleons. The
superscripts 1 and 2 label the nucleon.  These couplings  are
analytic functions of  $q^2$ for $q^2 < \Lambda_s^2$; the
propagator is
clearly analytic for $q^2 > \Lambda_s^2$.

In principle, we could consider a more sophisticated
model   consistent with the philosophy outlined above.  For
example,
one could explicitly include the exchange of three low-energy
pions
(with the quantum numbers of the rho) along with an omega
self-energy
due to its coupling with the three pion channel and a
longer-range part
of the $\omega$-N vertex due to three pion exchange.
In practice, one expects such effects to be small: in part they
serve to
simply widen the omega pole by an amount of no practical
significance
to the potential; other effects of coupling to the three pion
channel are small because they are weakly coupled.   In any
event, we will stick to the conventional assumptions underlying
meson exchange models and neglect such effects. In the remainder
of this paper we will ignore such effects.

 Let us now suppose that we had a detailed
microscopic model of
the $\omega$ meson which enables us to calculate a momentum
dependent $\omega$ self energy, $\pi_{\omega}(q^2)$.  As a matter
of
convention, we will include any
effects of  mass and wavefunction renormalizations of the
$\omega$  in
$\pi_{\omega}(q^2)$.   This means that $\pi_{\omega}$ and its
derivative
vanishes at $q^2=m_\omega^2$.
The omega exchange part of the N-N potential
with such a model is
given by
\begin{equation}
V_{\omega}(q^2) = \frac{( \hat{g}_{\omega}^{\rm
v}(q^2)\gamma^{(1)}_{\mu} \, +  \, \hat{g}_{\omega}^{\rm t}(q^2)
q^\alpha \sigma^{(1)}_{\alpha \mu}  ) (g^{\mu \nu} -
q^{\mu}q^{\nu}/m_{\omega}^2) (\hat{g}_{\omega}^{\rm v}(q^2)
\gamma^{(2)}_{\nu} + \hat{g}_{\omega}^{\rm t}(q^2) \sigma_{\nu
\beta}
q^\beta )  }{q^2 - m_\omega^2 + \pi_\omega(q^2)}\;.
\label{oep2}
\end{equation}
We have written the couplings as $\hat{g}_{\omega}^{\rm v,t}$ and
rather than $g_{\omega}^{\rm v,t}$ to make evident  the fact that
the
vertex functions used in the model in eq.~(\ref{oep2}) need not
be
the same
as the vertex functions used in the model in eq.~(\ref{oep}):
{\it these
vertex functions are phenomenological and depend on how the rest
of the problem is treated}.

Given that the vertex functions may differ between the two
models,  we note that the two models may be
identical---{\it i.e.} they may be two equivalent ways of
representing the same physics.   One way for this to occur is
if the vertex functions in the two models are related by
\begin{equation}
g_\omega^{\rm v,t} \, = \, \hat{g}_\omega^{\rm v,t} \, \left
(\frac{q^2
- m_\omega^2}{q^2 - m_\omega^2 + \pi_{\omega}(q^2)} \right
)^{1/2}.
\label{condition}
\end{equation}
Note
that
the square root factor is unity for $q^2=m_\rho^2$ due to the
renormalization of
$ \pi_{\omega}(q^2)$.
The  result (\ref{condition}) is not surprising in light of the
formal
analysis of Sect.II.  Neither the propagator off-shell
nor the vertex function are separately meaningful.

Given that vertex functions
are fit to some set of data, the only reason the condition in
eq.~(\ref{condition}) would not be satisfied would be due to
practical
and philosophical limitations
in the forms used in the fitting of the vertex functions.
The practical limitation is
that
one must take some limited  trial form for the phenomenological
coupling.   To the extent that meson exchange models make sense
in the regime where they are used, the trial forms must be rich
enough
to describe the data with reasonable precision.  Thus, apart from
the philosophical concerns discussed below, eq.~(\ref{condition})
can
be satisfied well enough so that any difference between the
potentials
of eqs. (\ref{oep}) and (\ref{oep2})
will have a small effect on the physics.
 The philosophical limitation
is that the vertex functions are supposed to only contain effect
of a
range shorter than $\Lambda_s^{-1}$.  Longer range effects are to
be included by explicit dynamics of the lighter
 degrees of freedom in the  problem.

Thus, the issue of whether the two models are equivalent comes
down to
whether both $g_{\rm v,t}(q^2)$ and $\hat{g}_{\rm v,t}(q^2)$ can
be analytic for $q^2<\Lambda_s^2$ while satisfying
eq.~(\ref{condition}).   In effect, the question is whether
\begin{equation}
f(q^2) =  \left (\frac{q^2 - m_\omega^2}{q^2 - m_\omega^2 +
\pi_{\omega}(q^2)} \right )^{1/2}
\end{equation}
is analytic for $q^2 < \Lambda_s^2$.   Non-analyticity can occur
when either
the numerator or denominator vanishes or when $\pi_{\omega}$ is
non-analytic.   In fact, we should relax this restriction
slightly---the
non-analyticity associated with the $\omega$ coupling to three
low
energy pions which slightly broadens the pole and gives a small
non-resonant contribution is, as discussed above,
innocuous.  In any event, this issue does not arise in the
context of the
models in refs. \cite{GHT92}-\cite{MTRC94}.

Clearly, the analytic structure of $f(q^2)$ depends in detail on
the
choice of model.    The simplest way to make the physics explicit
is to
make a spectral representation\cite{spectralref} for the
propagator:
\begin{equation}
\frac{1}{q^2 - m_\omega^2 + \pi_{\omega} (q^2)} = \int \, {\rm
d}s \,
\frac{\rho(s) }{q^2 - s + i \epsilon} + {\rm subtraction \quad
terms}.
 \label{prop}
\end{equation}
Different models will give rise to different spectral
functions.  However,  if the model is realistic, the
only substantial spectral strength for $q^2< \Lambda_s^2$ occurs
at
or near the
omega pole.
Accordingly any model
which gives significant amounts of  spectral strength below
$\Lambda_s^2$
(apart from the $\omega$ pole), can be considered as
unrealistic  in our philosophy.
If, however,
all of the spectral strength is either at the $\omega$ pole or
above
$\Lambda_s^2$, then $f(q^2)$ is analytic for $q^2$ below
$\Lambda_s^2$.
The
apparent
 non-analyticity due to the denominator vanishing at $q^2=
m_\omega^2$ is
precisely canceled by a vanishing numerator. (Recall,
all renormalization effects are included in $\pi_\omega$ so that
the
position
of the $\omega$ pole does not shift.) In this  case, one may
re-define the vertex
functions according to eq.~(\ref{condition}).

To see how the spectral representation constrains the allowable
forms of the self energies consider  consider the following
simple example in which the self energy has the form

\begin{equation}
 \pi_{\omega} (q^2)=(q^2-m_\omega^2)^2 Bq^2. \label{exam}
\end{equation}
This form is motivated by
the renormalization requirements, that $\pi_\omega(q^2)$ and its
derivative vanish at $q^2=0$.
One determines the
nucleon-nucleon potential generated by the propagator of
Eq.(\ref{prop}) by taking $q^2$ to be space-like $q^2=-Q^2<0$;
the potential is proportional to
the integral $$\int  {\rm d}Q Q{\sin (Qr)\over r}
\frac{1}{-Q^2 - m_\omega^2 -(Q^2+m_\omega^2)^2 BQ^2 }.$$
One does the contour integration by identifying the poles. There
is always a pole at $Q^2=-m_\omega^2$ which is the standard term
expected from the exchange of an $\omega^2$-meson. There are
other poles at positions determined by the value of B. One finds
that if $ m_\omega^4>  m_\omega^4+4/B\ge0$ there will be poles
with $Q^2$ real and negative. At least one of the poles must be
at $|Q^2|<m_\omega^2$, which
is unrealistic in our philosophy. If
$m_\omega^4+4/B >  m_\omega^4 $, the poles occur for $Q^2>0$
which are physically un-allowable tachyonic excitations.
Similarly, if $m_\omega^4+4/B<0$, there are
poles off the real axis, which violates the spectral
representation and also renders the model  for $ \pi_{\omega}
(q^2)$ as useless.
This analysis demonstrates that a spectral function of the
form in Eq.(\ref{exam}) is not viable.

Let us now summarize the effects of  the momentum dependence of
the
$\omega$ self energy
on the meson  exchange potential.    In any realistic model,
({\it i.e.} any model without unphysical low $q^2$ spectral
strength in
the $\omega$  propagator) all
of the effects of the momentum dependence of the
self-energy can be re-absorbed into  momentum dependence of the
phenomenological vertex functions.  Accordingly, there are no
observable physical effects in the NN potential induced by such a
momentum
dependent  self-energy.    Moreover,
including the short range part of the momentum dependence in the
propagator of  a meson  exchange model
 violates the
bookkeeping arrangement in which all of the short
range effects are segregated into phenomenological
vertices.

\section{The Charge Symmetry Breaking NN Potential And The
Momentum
Dependence of $\rho -\omega$ Mixing}

The preceeding section gives us a paradigm for what happens in
the
charge symmetry breaking part
of the potential.  We will show,  for any realistic model of the
momentum dependence of the mixing amplitude, that all of the
effects of
the momentum dependence can be absorbed into phenomenological
short
ranged charge-symmetry-breaking nucleon-vector meson couplings.

Consider the charge-symmetry-breaking potential
arising from vector meson exchange.   Let us begin
by implementing this according to the philosophy
of scale separation discussed in the previous two sections.
Assuming that
only
the meson exchanges we need
to consider are the $\rho$ and $\omega$, the charge
symmetry breaking interaction potential can be written as
\begin{eqnarray}
V_{\omega,\rho}^{\rm CSB} (q^2) \, =
 \, \frac{[g_\omega^{\rm v}(q^2) \gamma_{\mu}^{(1)}  +
g_\omega^{\rm
t}(q^2) q^\alpha \sigma_{\alpha \mu}^{(1)}] [g^{\mu \nu} -
q^{\mu}q^{\nu}/m_{\omega}^2] [g_\omega^{\rm v \, CSB}(q^2)
\tau_3^{(2)}
\gamma_{\nu}^{(2)}  + g_\omega^{\rm t \, CSB}(q^2)  \tau_3^{(2)}
\sigma_{\nu \beta}^{(2)} q^\beta]  }{q^2 - m_\omega^2}  \nonumber
 \\ \nonumber \\
+ \, \frac{ [g_\omega^{\rm v}(q^2) \gamma_{\mu}^{(2)}  +
g_\omega^{\rm
t}(q^2) q^\alpha \sigma_{\alpha \mu}^{(2)}] [g^{\mu \nu} -
q^{\mu}q^{\nu}/m_{\omega}^2]  [g_\omega^{\rm v \, CSB}(q^2)
\tau_3^{(1)} \gamma_{\nu}^{(1)}  + g_\omega^{\rm t \, CSB}(q^2)
\tau_3^{(1)} \sigma_{\nu \beta}^{(1)} q^\beta]  }{q^2 -
m_\omega^2}
   \nonumber \\ \nonumber \\
 + \, \frac{
[g_\rho^{\rm v} (q^2)  \gamma_{\mu}^{(1)} \tau_3^{(1)}  +
g_\rho^{\rm
t}(q^2) q^\alpha \sigma_{\alpha \mu}^{(1)} \tau_3^{(1)}][g^{\mu
\nu} -
q^{\mu}q^{\nu}/m_{\rho}^2][g_\rho^{\rm v \, CSB}(q^2)
\gamma_{\nu}^{(2)}  + g_\rho^{\rm t \, CSB}(q^2)   \sigma_{\nu
\beta}^{(2)} q^\beta]  }{q^2 - m_\rho^2}  \nonumber \\ \nonumber
\\
 + \, \frac{
[g_\rho^{\rm v} (q^2) \tau_3^{(2)} \gamma_{\mu}^{(2)}   +
g_\rho^{\rm
t}(q^2) q^\alpha \sigma_{\alpha \mu}^{(2)} \tau_3^{(2)}] [g^{\mu
\nu} -
q^{\mu}q^{\nu}/m_{\rho}^2][g_\rho^{\rm v \, CSB}(q^2)
\gamma_{\nu}^{(1)}  + g_\rho^{\rm t \, CSB}(q^2)   \sigma_{\nu
\beta}^{(1)} q^\beta]  }{q^2 - m_\rho^2}  \nonumber \\ \nonumber
\\
\, + \, m^2_{\rho \omega} \, \, \frac{[g_\omega^{\rm v}(q^2)
\gamma_{\mu}^{(1)}  + g_\omega^{\rm t}(q^2) q^\alpha
\sigma_{\alpha \mu}^{(1)}]
[g^{\mu \gamma} - q^{\mu}q^{\gamma}/m_{\omega}^2]
 [g^{\gamma \nu} - q^{\gamma}q^{\nu}/m_{\rho}^2]
[g_\rho^{\rm v}(q^2) \tau_3^{(2)}\gamma_{\nu}^{(2)}  +
g_\rho^{\rm
t}(q^2)  \tau_3^{(2)} \sigma_{\nu \beta}^{(2)} q^\beta]
}{(q^2-m_\omega^2)(q^2-m_\rho^2)} \nonumber \\ \nonumber \\
+ \, m^2_{\rho \omega} \, \, \frac{
[g_\omega^{\rm v}(q^2) \gamma_{\mu}^{(2)}  + g_\omega^{\rm
t}(q^2)
q^\alpha \sigma_{\alpha \mu}^{(2)}]
[g^{\mu \gamma} - q^{\mu}q^{\gamma}/m_{\omega}^2]
 [g^{\gamma \nu} - q^{\gamma}q^{\nu}/m_{\rho}^2]
[g_\rho^{\rm v}(q^2) \tau_3^{(1)}\gamma_{\nu}^{(1)}  +
g_\rho^{\rm
t}(q^2)  \tau_3^{(1)} \sigma_{\nu \beta}^{(1)} q^\beta] }
{(q^2-m_\omega^2)(q^2-m_\rho^2)} \nonumber\ \\ \label{CSB1}
\end{eqnarray}
This form is rather general: in addition to $\rho-\omega$ mixing,
it
explicitly includes possible charge symmetry breaking couplings
between
the vector mesons and the nucleons arising from short
distance effects: these couplings are  labeled
by the superscript CSB.  The coefficient $m^2_{\rho \omega}$ is
the
mixing parameter which in this model is
taken to be independent of $q^2$.

It should be noted that the general form of eq.~(\ref{CSB1}) is
consistent  with the general philosophy of meson exchange
used here.
In particular all short-ranged effects are
merely
parameterized, while the long ranged effects are treated
dynamically in terms of the mesons.  It is for this reason, that
we
must include the $\rho-\omega$ mixing
explicitly rather than including all of the effects in terms
of the charge symmetry breaking couplings.

The couplings $g_{\omega,\rho}^{\rm v,t  } (q^2)$
are presumed to have been determined in fits to the
charge symmetry conserving interactions.
In principle, the coupling constants $g_{\omega,\rho}^{\rm v,t
\, CSB}
(q^2)$ must be determined phenomenologically from experimental
data on
charge symmetry breaking.  In fact, in the treatments of CSB in
refs.
\cite{CB87,BI87,M94,MVO94} these couplings were
all taken to be zero.
In that work, model assumptions and existing nucleon-nucleon
and pion-nucleon scattering data were used to
make {\it a priori} arguments that these couplings should be
small and
hence could be neglected.  See, for example, Refs. \cite{MNS90}
and \cite{MVO94}
which reviews the charge-dependence of the couplings.
 The neglect of charge dependence in the meson-nucleon coupling
constants is not
invalidated by present data.
In particular,
descriptions of all known  CSB effects do
not require the inclusion  of such terms.
Had the data required the
inclusion of such terms they could have been included without
violating
the spirit of a meson
exchange potential model.

Now suppose, we had a detailed model for the
structure of the vector mesons  in which the $\rho-\omega$
mixing  amplitude
has a nontrivial momentum dependence.
The form for the CSB potential is very similar to the form above:
\begin{eqnarray}
V_{\omega,\rho}^{\rm CSB} (q^2)  =
  \frac{[g_\omega^{\rm v}(q^2) \gamma_{\mu}^{(1)}  +
g_\omega^{\rm
t}(q^2) q^\alpha \sigma_{\alpha \mu}^{(1)}] [g^{\mu \nu} -
q^{\mu}q^{\nu}/m_{\omega}^2] [\hat{g}_\omega^{\rm v \, CSB}(q^2)
\tau_3^{(2)} \gamma_{\nu}^{(2)}  + \hat{g}_\omega^{\rm t \,
CSB}(q^2)
\tau_3^{(2)} \sigma_{\nu \beta}^{(2)} q^\beta]  }{q^2 -
m_\omega^2}
\nonumber  \\ \nonumber \\
+ \frac{ [g_\omega^{\rm v}(q^2) \gamma_{\mu}^{(2)}  +
g_\omega^{\rm
t}(q^2) q^\alpha \sigma_{\alpha \mu}^{(2)}] [g^{\mu \nu} -
q^{\mu}q^{\nu}/m_{\omega}^2]  [\hat{g}_\omega^{\rm v \, CSB}(q^2)
\tau_3^{(1)} \gamma_{\nu}^{(1)}  + \hat{g}_\omega^{\rm t \,
CSB}(q^2)
\tau_3^{(1)} \sigma_{\nu \beta}^{(1)} q^\beta]  }{q^2 -
m_\omega^2}
   \nonumber \\ \nonumber \\
 +  \frac{
[g_\rho^{\rm v} (q^2)  \gamma_{\mu}^{(1)} \tau_3^{(1)}  +
g_\rho^{\rm
t}(q^2) q^\alpha \sigma_{\alpha \mu}^{(1)} \tau_3^{(1)}][g^{\mu
\nu} -
q^{\mu}q^{\nu}/m_{\rho}^2][\hat{g}_\rho^{\rm v \, CSB}(q^2)
\gamma_{\nu}^{(2)}  + \hat{g}  _\rho^{\rm t \, CSB}(q^2)
\sigma_{\nu
\beta}^{(2)} q^\beta]  }{q^2 - m_\rho^2}  \nonumber \\ \nonumber
\\
 +  \frac{
[g_\rho^{\rm v} (q^2) \tau_3^{(2)} \gamma_{\mu}^{(2)}   +
g_\rho^{\rm
t}(q^2) q^\alpha \sigma_{\alpha \mu}^{(2)} \tau_3^{(2)}] [g^{\mu
\nu} -
q^{\mu}q^{\nu}/m_{\rho}^2][\hat{g}_\rho^{\rm v \, CSB}(q^2)
\gamma_{\nu}^{(1)}  + \hat{g}_\rho^{\rm t \, CSB}(q^2)
\sigma_{\nu
\beta}^{(1)} q^\beta]  }{q^2 - m_\rho^2}  \nonumber \\ \nonumber
\\
 + \, m^2_{\rho \omega}(q^2) \, \, \frac{[g_\omega^{\rm v}(q^2)
\gamma_{\mu}^{(1)}  + g_\omega^{\rm t}(q^2) q^\alpha
\sigma_{\alpha \mu}^{(1)}]
[g^{\mu \gamma} - q^{\mu}q^{\gamma}/m_{\omega}^2]
 [g^{\gamma \nu} - q^{\gamma}q^{\nu}/m_{\rho}^2]
[g_\rho^{\rm v}(q^2) \tau_3^{(2)}\gamma_{\nu}^{(2)}  +
g_\rho^{\rm
t}(q^2)  \tau_3^{(2)} \sigma_{\nu \beta}^{(2)} q^\beta]
}{(q^2-m_\omega^2)(q^2-m_\rho^2)} \nonumber \\ \nonumber \\
+  m^2_{\rho \omega}(q^2) \, \, \frac{
[g_\omega^{\rm v}(q^2) \gamma_{\mu}^{(2)}  + g_\omega^{\rm
t}(q^2)
q^\alpha \sigma_{\alpha \mu}^{(2)}]
[g^{\mu \gamma} - q^{\mu}q^{\gamma}/m_{\omega}^2]
 [g^{\gamma \nu} - q^{\gamma}q^{\nu}/m_{\rho}^2]
[g_\rho^{\rm v}(q^2) \tau_3^{(1)}\gamma_{\nu}^{(1)}  +
g_\rho^{\rm
t}(q^2)  \tau_3^{(1)} \sigma_{\nu \beta}^{(1)} q^\beta] }
{(q^2-m_\omega^2)(q^2-m_\rho^2)}\nonumber
\end{eqnarray}
\begin{equation}
\label{CSB2}
\end{equation}
We have labeled the CSB coupling as
$\hat{g}_{\omega,\rho}^{\rm v,t \,CSB}$ rather than
$g_{\omega,\rho}^{\rm v,t \, CSB}$ to make explicit
the fact the CSB couplings in eq. (\ref{CSB2}) may be different
from
the CSB couplings in eq. (\ref{CSB1}).

The question we wish to address is whether the model
in eq. (\ref{CSB2}) is equivalent to the model in eq.
(\ref{CSB1}).
The issue comes down to whether
the effects of the momentum dependence of the mixing can be
entirely
absorbed into differences between  $\hat{g}_{\omega,\rho}^{\rm
v,t \,
CSB}$ and $g_{\omega,\rho}^{\rm v,t \, CSB}$ without introducing
any unnaturally long range effects into the CSB couplings.  We
shall show
that this can be done.

The $\rho-\omega$ mixing is measured rather accurately at the
pole at $q^2=m_\omega^2$.  Accordingly, it is
sensible to express
\begin{equation}
m^2_{\rho \omega}(q^2) = m^2_{\rho \omega} + \delta m^2_{\rho
\omega}(q^2)
\end{equation}
with  $\delta m^2_{\rho \omega}(m_{\omega}^2) = 0$.
Thus, the expression
$$\frac{ \delta m^2_{\rho
\omega}(q^2)}{(q^2-m_\rho^2)(q^2-m_\omega^2)} $$
has no $\omega$ pole.  All effects with this term are
indistinguishable from terms arising due $\rho$ exchange with a
CSB
vertex.  In particular, if
\begin{eqnarray}
\hat{g}_{\omega}^{\rm v,t CSB} & = & g_{\omega}^{\rm v,t CSB}
\label{CSB3}\\
\hat{g}_{\rho}^{\rm v,t CSB} & = & g_{\rho}^{\rm v,t CSB} -
\frac{
\delta m^2_{\rho \omega}(q^2)}{q^2-m_\omega^2} g_{\rho}^{\rm v,t}
\label{CSB4}
\end{eqnarray}
then the potential in eq. (\ref{CSB2}) is identical with the one
of
eq. (\ref{CSB1}).
This result can also be obtained from Feynman diagrams. Let an
$\omega$
be emitted from
a nucleon and then be converted via $m^2_{\rho \omega}(q^2)$ into
a $\rho$. One can draw a box which includes the strong vertex
and $m^2_{\rho \omega}(q^2)$. This box is the charge-dependent
$\rho$-nucleon coupling constant. Alternatively one can regard
the $m^2_{\rho \omega}(q^2)$ as part of the propagator. Either
way, the result is the same.

We can do a specific calculation.
For example, suppose  $\delta m^2_{\rho
\omega}(q^2)=m^2_{\rho\omega}/m_\omega^2
(q^2-m_\omega^2)$. This is a good approximation to the $m^2_{\rho
\omega}(q^2)$ obtained in the sum rule work of
Ref.\cite{HHKM94}. Then
the difference between $\hat{g}_{\omega}^{\rm v,t CSB}$ and
$ g_{\rho}^{\rm v,t CSB}$ is a simple constant
$\approx -.008g_\rho^{v,t}$; if
$ \hat{g}_{\rho}^{\rm v,t CSB}$ were chosen as the negative of
that constant, one
would obtain the standard form of the $\rho-\omega$ mixing
contribution to
the NN potential. See also Ref.\cite{G95}.

Moreover, for any reasonable model of the momentum
dependence of the mixing, eq.~(\ref{CSB4}) can be
satisfied without introducing unnaturally long ranged effects
into the
meson-nucleon vertex functions.  The
issues are completely analogous to  the ones raised in connection
with
the  $\omega$ exchange potential discussed in the previous
section.  First,
it should be noted that
there is no $\omega$ pole singularity on the right
hand side of  eq.~(\ref{CSB4})---it is eliminated  because
$\delta
m_{\rho \omega}^2$ vanishes at the $\omega$ pole.  Thus, the only
source of long range contamination of the couplings is in $\delta
m_{\rho \omega}^2 (q^2)$ itself.  Note, that by construction
$\delta
m_{\rho \omega}^2 (q^2)$ cannot have a singularity associated
with
either the $\rho$ or the $\omega$.
Moreover, we know that the only substantial strength in the
vector
channels at  $q^2 < \Lambda_s^2$ is through the
$\rho$ and $\omega$ mesons.  Thus, any model
which yields long range effects in $\delta m_{\rho \omega}^2
(q^2)$
must be regarded as unrealistic according to our philosophy.

\section{Summary}

We are working in the framework of boson exchange potentials.
This means that in realistic boson-exchange models
long range effects are included via boson
exchanges and that short range effects are included in the vertex
functions.
For any such realistic model
of the momentum dependence of the $\rho-\omega$ mixing parameter,
there are no effects in the CSB breaking potential
which cannot be absorbed into a redefinition
of  a short ranged CSB $\rho$-N vertex.  Thus,  a model which
provides
knowledge of the momentum dependence of the mixing parameter
alone, without
simultaneously giving a self-consistent model for
the short ranged CSB vector meson-nucleon couplings gives no
information about the CSB N-N
potential.

The work of refs.
\cite{GHT92}-\cite{MTRC94}
found major
differences between the CSB
potentials based on the on-shell $\rho-\omega$ mixing and models
with a
large momentum dependence. Our boson exchange model view is that
this is because
the short ranged CSB
vector
meson-nucleon coupling are assumed to be zero---
just as in the
models based on the
on-shell $\rho-\omega$ mixing.  However,
there is no reason, {\it a priori} that this assumption is
be true for the models under discussion.
  Indeed, there is {\it a posteriori} evidence that the
assumption may  be wrong:  the models based on
the on-shell $\rho-\omega$ mixing and negligible
$g_{\rho \omega}^{\rm v,t \, CSB}$  reproduce the available data
with reasonable accuracy.

\acknowledgements
One of us (TDC) thanks the Department of Physics and the
Institute of
Nuclear Theory at the University Of Washington for its
hospitality.
The authors thank the U.S. Department of Energy for
supporting this work;  TDC also acknowledges the
financial support of the National Science Foundation.

\end{document}